# Is a rational explanation of wave-particle duality possible?

S.A. Rashkovskiy

*Institute for Problems in Mechanics, Russian Academy of Sciences, Vernadskogo Ave., 101/1 Moscow, 119526, Russia, Tel. +7 495 5504647, E-mail: rash@ipmnet.ru*

**Abstract**

Wave-particle duality is one of the fundamental properties of matter and at the same time, one of the mysteries of modern physics. In this paper, we propose and analyze a new interpretation of the wave-particle duality, and propose a new rule that relates the corpuscular and wave properties of quantum particles, and replaces the Born's rule. We introduce a new representation about quantum particles as temporary objects that are born permanently in a wave field and spontaneously decayed due to their "instability". Thus, we fill with the physical meaning the standard notions of "creation" and "annihilation" of the particles, used in the second quantization formalism. We show that the traditional probabilistic interpretation of quantum mechanics in the form of the Born's rule is an approximate consequence of the proposed theory. The proposed theory allows explaining the indistinguishability of quantum particles and shows meaninglessness of the concept of their trajectories. We give an explanation of the Heisenberg uncertainty principle within the limits of this theory in terms of classical physics that allows us to return to an objective understanding of the measurement process and do not consider it as a compulsory part of the physical process.

## I. INTRODUCTION

Wave-particle duality is one of the fundamental properties of matter and also one of the mysteries of modern physics.

It was first introduced by A. Einstein in 1905 [1], who showed that when light is treated as a flux of particles (quanta) with energy $\hbar\omega$ that interact with the particles of a substance according to classical mechanics, one can explain all known laws of photoelectric phenomena. The discovery of the Compton effect [2] became a triumph of the corpuscular hypothesis, which was easily explained by treating light as a flux of Einstein's quanta that possess energy $\hbar\omega$ and momentum $\hbar\omega/c$. Later, it became clear that other interactions of light with matter (the photochemical effect, fluorescence, etc.) have simple explanations if light is considered to be a flux of quanta, but are difficult to explain if light is considered to be a continuous wave. Conversely, the interference and diffraction phenomena cannot be explained if light is considered to be a flux of particles (quanta), but are naturally described within the limits of the classical wave theory. Thus, the entire set of experimental evidence indicates that in some experiments, light behaves like a wave, while in others, light behaves as a flux of quanta, or photons.

Precision experiments [3] have shown that the wave properties of light cannot be considered a property of the photon flux (which could be interpreted as a result of interactions of the particles in the flux); it was found that each individual photon should possess wave properties and can undergo interference.

Discovery of the wave-particle duality formulated some fundamental problems, the most important of which were the following: (i) how can a type of matter (light) possess such incompatible (in terms of classical physics) properties, specifically both wave and particle flux properties? (ii) what is a photon? These problems were posed by A. Einstein himself, who made several attempts to answer them [4-6], but was not able to do so.

On 12 December 1951, A. Einstein wrote to M. Besso: "*All these 50 years of conscious brooding have brought me no nearer to the answer to the question: What are light quanta?*"

After more than 100 years since the introduction of the wave-particle duality into physics, we can repeat A. Einstein's statement that we are not able to answer the question: what is a photon?

Modern quantum theory provides a formal answer to these questions, abandoning the classic images of waves and particles. Compatibility between the wave and corpuscular properties of light in quantum theory is achieved using a probabilistic interpretation of optical phenomena that can be simply described as follows: "the probability of photon localization at some point is proportional to the intensity of the light wave at this point, calculated on the basis of the methods of wave optics" [7]

$$n \sim \langle \mathbf{E}^2 \rangle \qquad (1)$$

where $n$ is the "concentration" of the photons at a given point, $\mathbf{E}$ is the strength of the electric field of the light wave at this point and $\langle ... \rangle$ indicates an averaging over time.

The idea of wave-particle duality provided a significant boost to all of physics in the early 20th century.

In 1924, Louis de Broglie generalized the wave-particle duality to all forms of matter [8]; it turned out that not only light, but also ordinary matter, has both corpuscular and wave properties. Later, it was experimentally demonstrated in a series of magnificent and ingenious experiments that a variety of particles (electrons, protons, neutrons, atoms, and molecules) possess wave properties [9-15]; moreover, these properties could not be ascribed to a flux of particles, but rather to individual particles [10]. Very recently, it was found that even macroscopic particles can demonstrate wave-particle duality [16]. However, the latter, in our opinion, should be considered a classical analogy, rather than a possible interpretation of quantum laws in the microcosm.

With the advent of quantum mechanics, the problem of interpreting the wave-particle duality has been additionally exacerbated. We note that the interpretation of quantum mechanics, in the end, is reduced to attempts to explain the wave-particle duality. One can say that the different interpretations of quantum mechanics are, in fact, the different explanations of the wave-particle duality.

A formal (canonical for modern physics) explanation of the wave-particle duality is given by the Copenhagen interpretation of quantum mechanics [7], which is based on (i) Born's probabilistic interpretation [17], which, in fact, generalizes the probabilistic interpretation for photons (1) to any kind of matter; (ii) the Heisenberg uncertainty principle [18], and (iii) Bohr's complementarity principle [19], which states that in some experiments, a quantum object behaves like a wave, while in others as a particle, and these properties never appear simultaneously. The Copenhagen interpretation is agnostic (irrational) because it states that, in principle, it is impossible to imagine these objects that we call quantum particles (photons, electrons, etc.). It proposes that our knowledge, which is based on the classical images obtained in the macrocosm, is limited and simply unable to accept that which belongs to the microworld [7].

Whilst the Copenhagen interpretation is shared by many modern physicists [20], it is not universally accepted. Despite its undeniable assistance in the formal interpretation of quantum phenomena (and especially the results of mathematical theory), the Copenhagen interpretation causes dissatisfaction on the part of physicists. This explains the incessant efforts to find a more reasonable interpretation of the wave-particle duality, one that is compatible with the classical images and classical way of thinking. Such an interpretation, if possible in principle, could be called a rational one (contrary to the irrational Copenhagen interpretation) because it would only be enacted using the classic pictorial images.

Attempts to give a rational explanation for quantum phenomena have resulted in a number of witty and more or less fantastic interpretations of quantum mechanics [20-26].

However, we can say that at present, there is no consistent rational interpretation of the wave-particle duality, and, as a consequence, all of quantum mechanics. Actually, at present, *we have a well-composed mathematical theory; however, we cannot imagine what object it actually describes*.

It is a quite common view that a rational interpretation of quantum mechanics is impossible in principle.

In this paper, we will attempt to refute this opinion.

We show that not all the possible interpretations of the wave-particle duality have been considered yet. In particular, we propose a new interpretation, which we call "a rational interpretation", based on clear classical concepts of waves and particles; however, the link between these quantum object properties is new.

## II. PHOTONS

In this interpretation, we will proceed from the *assumption that photons exist objectively*.

Both corpuscular and wave properties of light can be observed by projecting the diffraction pattern on a photographic plate. A sequence of dark and light fringes observed on the photographic plate, under the microscope, is composed of discrete points, which is interpreted as the point of impact of single photons on a photographic plate. Thus, the dark and light fringes are the areas on which, respectively, large or small number of photons impinge. The observed diffraction pattern is in complete agreement with expression (1). This indicates that the photons are distributed non-uniformly in the diffracted light beam. If one shines a plane light wave on the same screen of photosensitive material, its action at all points of the screen will be the same, causing a uniform blackening of the photographic plates. In other words, the plane waves of light can be represented as a uniform flux of photons (at least on the average for the time of exposure). Thus, we can conclude that the diffraction and interference of a plane light wave leads to the redistribution of photons in the light beam; as a result, the distribution of photons in space becomes non-uniform.

The reason for the non-uniform redistribution of photons in the light beam due to interference and diffraction cannot be explained using the concept of the photon as an unchanging classical particle.

The double-slit experiment [27] is traditionally used for directly demonstrating wave-particle duality. This experiment is considered to be a test of new interpretations of the wave-particle duality; any interpretation that fails to explain the double-slit experiment does not merit notice.

However, we believe that the Wiener experiments [28] are more suitable for this purpose. These experiments are a direct demonstration of the redistribution of photons in space due to the interference of plane light waves.

All known alternative interpretations of the wave-particle duality [20-26] explain the double-slit experiment in some way; however, they often find themselves powerless to explain the Wiener experiments.

Wiener [28] investigated the field of a standing light wave that was formed by reflection of an incident plane wave from a mirror. A thin photographic emulsion film, which was located at a very small angle relative to the reflecting surface, was placed in the field of the standing light wave. In the first experiment, Wiener investigated the normal incidence of the light beam on the mirror. After developing the film, black equidistant fringes and transparent intervals between the fringes were found. Analysis of the optical field in these experiments and comparison with the pattern on the photographic film shows that the black fringes correspond to the areas in the vicinity of the maxima of the electric field $\mathbf{E}^2$, in complete agreement with interpretation (1). If we remove the mirror (i.e., acting on the same film only by a single incident plane wave), the blackening of the film, as expected, is uniform. Thus, the interaction of the incident and reflected light beams leads to a redistribution of photons in space; the uniform distribution of the photons in the incident and reflected waves becomes significantly non-uniform in the standing wave.

Similar experiments were performed with fluorescent [29] and photoemission films [30], which were used as standing wave detectors, instead of the photographic emulsion used by Wiener. In these cases, the maximum effect was also found in the antinodes of the electric field, which from the corpuscular perspective means that in these areas there is an "enhanced concentration of photons".

In the second experiment, Wiener [28] used his setup to investigate interference phenomena in linearly polarized light at an incidence angle of 45°. He found that if the direction of electrical oscillations in the incident light was perpendicular to the incidence plane, the dark areas in the emulsion form a set of equidistant parallel fringes; this corresponds to essentially a non-uniform distribution of photons in space. However, if the vector of electrical oscillations in the incident light lay in the plane of incidence, the blackening of the film was uniform, and consequently, the photons in space were distributed uniformly. All these results are in complete agreement with the wave theory [31] and with the traditional quantum interpretation (1). However, attempts to explain the non-uniform redistribution of photons in space in these experiments, within the limits of the concept of photons as classical particles, have been unsuccessful. One can state that all attempts to combine the wave and particle properties of light based on interpretation (1) have failed.

In our opinion, the probabilistic interpretation based directly on expression (1) is precisely the reason for the conflict in physics, which we call the wave-particle duality.

We show that the rejection of the probabilistic interpretation of experimental facts in the form of (1) and its replacement by a more general relation, as well as the abandonment of the idea of the photon as an unchanging particle, allows one to resolve this logical conflict without contradiction and to use classic visual images to explain the quantum phenomena.

Let us make the following basic assumptions:

1. *A wave is primary*; i.e., only a continuous field that can be described by the Maxwell equations (at least in classical experiments with interference) exists.

2. *Photons are temporary formations* that are periodically born and perish in space, and the processes of their creation and destruction are determined by the total wave field resulting from the interference of different waves at a given point and time.

3. The rate of creation of photons at a given point depends on the intensity of the total electric field of the wave at that point:

$$v_+ = \gamma \mathbf{E}^2 \tag{2}$$

where $v_+$ is the number of photons generated per unit volume per unit time and $\gamma$ is a constant.

4. *Photons are unstable formations* that have a mean lifetime $\tau$ and spontaneously decay independently of each other. The rate of disappearance of photons $v_-$ (the number of photons that disappear per unit volume per unit time) can be described by a relation similar to that for radioactive decay

$$v_- = \frac{1}{\tau} n \tag{3}$$

We note that this does not violate the energy conservation law because it is assumed that the energy required for photon creation is taken from the wave field $\mathbf{E}$ (one can say that the field gives birth to photons) and it returns to the wave field after the spontaneous decay of the photon. Therefore, for photon concentration $n$, one can write the kinetic equation

$$\frac{dn}{dt} = v_+ - v_- \tag{4}$$

which, when taking into account expressions (2) and (3), becomes

$$\frac{dn}{dt} = \gamma \mathbf{E}^2 - \frac{1}{\tau} n \tag{5}$$

Equation (5) has the solution

$$n = \gamma \exp\left(-\frac{t}{\tau}\right) \int_0^t \mathbf{E}^2(t') \exp\left(\frac{t'}{\tau}\right) dt' \tag{6}$$

Using (6), it is easy to calculate the mean photon concentration at a given point:

$$\langle n \rangle = \gamma \tau \langle \mathbf{E}^2 \rangle \tag{7}$$

Expression (7) exactly coincides with relation (1), which is the basis for the probabilistic interpretation of quantum mechanics. However, as we see, it is only a consequence of the more general principle stated above and described by equation (5).

According to the modern concept of quantum mechanics, all the energy of light is transported by photons. Without discussing this statement in detail, we will use it as a basis. In this case, we must then write

$$\hbar \omega \langle n \rangle = \frac{\langle \mathbf{E}^2 \rangle}{4\pi} \tag{8}$$

Comparing (7) and (8), one obtains

$$\frac{1}{\gamma\tau} = 4\pi\hbar\omega \tag{9}$$

Thus, equation (5) can be written in the form

$$\frac{dn}{dt} = \gamma\left(\mathbf{E}^2 - 4\pi\hbar\omega n\right) \tag{10}$$

It follows that for large wave intensities $\mathbf{E}^2$, one can describe the photon concentration, which corresponds to the classical limit of quantum theory [32]. At low wave intensities $\mathbf{E}^2$, one can only relate the probability to detect a photon at a given point. In this case, expressions (2) - (5) retain their forms; however, $n$ should be considered the probability density of finding a photon at a given point, while $v_+$ and $v_-$ are, respectively, the probabilities of creation and destruction of photons per unit time per unit volume.

The fundamental point of this theory is the permanently occurring process of formation (birth) of photons from the wave field and their spontaneous decay. It allows the wave and particle properties of light to be combined consistently in a single theory. In fact, due to the interference of waves, a redistribution of the intensity of radiation in space occurs in accordance to classical electrodynamics (optics). Simultaneously, due to the destruction and creation of photons, a redistribution of the "photon field" occurs in space; the distribution of photons in space almost instantly adjusts to the new distribution of the wave field intensity.

We note that the creation and destruction of the photons are by themselves not new concepts in quantum physics. In fact, the second quantization formalism with its creation and annihilation operators initially acts via the "creation" and "annihilation" of particles and considers all processes (including the motion of particles in space) as the annihilation of particles in one state and their birth in another. A new theory has proposed that photons are created by the wave field and that the rate of their creation (2) depends on the intensity of the total wave field at a point, which can (at least approximately) be calculated according to the laws of classical electrodynamics or wave optics.

Let us consider what consequences follow from the proposed theory.

We argue that *if photons really exist*, they should be considered temporary formations that randomly appear at different points in space and spontaneously decay over a time $\tau$. This means that we cannot trace a single photon, and, moreover, cannot say that a photon detected at a given point, for example, after interacting with an electron, is the same photon that was previously in another point in space and moved to this point. Thus, it is not sensible to talk about the trajectory of a photon. As known from quantum mechanics, the rejection of the concept of the trajectory of particles (photons) is necessary to avoid controversy in the interpretation of experimental evidence. The proposed theory shows that the concept of photon trajectories in principle has no meaning and thus provides a natural justification of this postulate of quantum theory. For the same reason, we can say that all photons are indistinguishable particles. The indistinguishability of particles (photons) in quantum mechanics is also introduced as a postulate to avoid conflict with the experimental data; in our theory, the indistinguishability of particles is a natural consequence. In fact, it is impossible to understand the indistinguishability of particles that are treated as unchanging classical corpuscles. However, if the photons are temporary formations, in principle, we cannot mark them in any manner and cannot say that a given photon is a "successor" of any other photon that had previously existed at another point in space.

Let us consider how one can understand the statement that photons are created and perish periodically in space and that the processes of creation and annihilation are determined by the total wave field at a given point. This can be easily interpreted if the photons are considered clots of the field. To do this, we must assume that the actual light field (i.e., the electromagnetic field) is a classical nonlinear field and that uniform monochromatic waves of such a field are absolutely unstable; small disturbances of the field, which always exist, begin to grow indefinitely, forming a compact area with high field strength. The field strength in these clots can significantly exceed the mean intensity of the (linear) field. If a particle (e.g., an electron) arrives at these areas, the action on it from the field will be pulsed and stronger than that of the mean field; in experiments, this is considered a collision of an electron with a quantum of the wave field, i.e., with a photon. After reaching the maximum amplitude, the clot begins to "dissipate" or decay due to its instability, which would be perceived as spontaneous decay of a photon. The number of the most unstable perturbations (i.e., the perturbations with the highest growth rate, which will survive after competition with other field disturbances) per unit volume should be approximately $\sim \mathbf{E}^2$. This means that the critical wavelength $\Lambda$ of unstable field disturbances must satisfy the condition $\langle \mathbf{E}^2 \rangle \Lambda^3 \sim \hbar\omega$. We assume that the "germs" of photons (critical disturbances) have approximately the same sizes in all directions, i.e., are isotropic.

We note that this scheme is not particularly unique; nonlinear waves that have similar properties are well known in physics. For example, in nonlinear optics, there exists a filamentation phenomenon - the collapse of an intense laser beam due to the nonlinear Kerr effect [33-35]. Under certain conditions, so-called multiple filamentation occurs [34,35], resulting in a laser beam divided into multiple clots, or filaments (sometimes called light bullets), the energy density of which is many orders of magnitude greater than the energy density of the initial (linear) beam. Each filament transports approximately the same amount of energy. These clots (filaments) are unstable formations and dissipate quickly. The creation-destruction process of filaments is repeated. As a result, the laser beam after filamentation is not a smooth continuous wave, but rather a flux of energy clots (filaments) with an extremely large (compared to a linear optical wave) energy density that are permanently created and annihilated due to the instability of the nonlinear field. Here, we can see a direct analogy between the phenomena of multiple filamentation of an intense laser beam and the process described above of the formation and decay of photons in a hypothetical non-linear electromagnetic field. Of course, we cannot completely identify photons with filaments, as the latter have (i) a different nature (filamentation occurs only in a nonlinear optical medium - gases and liquids - while photons exist even in vacuum), (ii) different parameters (the filament sizes are significantly larger than the length of the carrier wave, while in the classical limit of quantum electrodynamics, [32] "the size of the photon" should be much smaller than the wavelength), and (iii) different behaviour (multiple filamentation occurs when increasing the intensity of the laser beam above the critical value, while the quantum properties of light appear more strongly for less intensity, i.e., approaching the linear limit - Maxwell's equations - corresponds not to low intensity of light, but rather high intensity). However, the example above shows that the well-known nonlinear phenomena in the electromagnetic waves are quite consistent with the rational interpretation of the wave-particle duality of light developed in this paper.

### III. NON-RELATIVISTIC MATTER

The theory developed above can easily be extended to non-relativistic matter.

In non-relativistic quantum mechanics, particles (electrons, protons, neutrons, etc.) are described by a complex scalar wave function $\psi(t,\mathbf{r})$, which obeys the Schrödinger equation. Thus, the Schrödinger equation describes the continuous field $\psi(t,\mathbf{r})$ that reflects the wave properties of quantum particles, similar to how the electric and magnetic fields reflect the wave properties of light. The relationship between the wave and corpuscular properties of quantum particles is established by Born's rule [17]:

$$p(t,\mathbf{r}) = |\psi(t,\mathbf{r})|^2 \tag{11}$$

where $p(t,\mathbf{r})$ is the probability density of finding the particle at a given point.

Born's rule (11) is a direct generalization of rule (1) relating the wave and particle properties of light and represents a mathematical expression of the wave-particle duality of non-relativistic quantum particles.

Thus, quantum mechanics, in fact, consists of two independent parts: the wave equation (e.g., the Schrödinger equation), which describes a continuous field $\psi(t,\mathbf{r})$ reflecting the wave properties of particles, and Born's rule (11) reflecting the corpuscular behaviour of particles.

Born's rule is an independent postulate that does not follow from the wave equation. There have been attempts to derive Born's rule from within standard quantum mechanics. For example, W. H. Zurek presented a derivation of Born's rule based on a mechanism termed "environment-assisted invariance" [36-38]. However, it should be noted that this derivation cannot be considered physical. It is rather an attempt to find a class of systems for which Born's rule is obeyed.

It is easy to see that the probabilistic interpretation of the wave function, based directly on Born's rule (11), is precisely the reason for the conflict in physics called the wave-particle duality. Despite numerous attempts to date, all have failed to construct a theory that, using the classical images of wave and particle, allows one to obtain Born's rule (11) in a non-contradictory manner. All attempts to combine the wave and particle properties of the particles on the basis of rule (11) have failed. It is for this reason that the complementarity principle [19] became one of the guiding principles in quantum mechanics.

By analogy with photons, the wave-particle duality of non-relativistic particles can be explained if we abandon the image of quantum particles as unchanging point objects that permanently move in space.

Let us make the following basic assumptions:

1. *The wave is primary*. This means that we consider the continuous field $\psi(t,\mathbf{r})$ that can be described by the wave equation, e.g., by the Schrödinger equation, as a real physical field, like the electromagnetic field. In our opinion, the complex wave function $\psi(t,\mathbf{r})$ is a simplified form of the description of an actual physical field. We assume that the frequency of this wave is related to the rest mass of the particle by $\omega = mc^2/\hbar$ and significantly higher than the frequency $E/\hbar$, where $E$ is the non-relativistic energy of the particle.

2. *The particles (electrons, protons, etc.) are temporary formations* that are periodically created and annihilated in space, and the processes of their creation and destruction are determined by the total wave field $\psi(t,\mathbf{r})$ at a given point and time, i.e., by the field resulting from the interference of different waves.

3. The rate of particle creation at a given point depends on the intensity of the total wave field at this point, $\psi(t,\mathbf{r})$.

The fundamental point of this theory is the postulate that the field can create particles. Particle creation is a random process, most likely associated with a nonlinearity of the real field and with the disturbances that always exist in space. For this reason, the process of particle creation is characterized by a probability of creation during the time $dt$ in the volume $d\mathbf{r}$: $\nu_+(t,\mathbf{r})d\mathbf{r}dt$. Generalizing expression (2), we can postulate:

$$v_+ = \gamma\beta \frac{|\psi(t,\mathbf{r})|^2}{\hbar} \tag{12}$$

where $\beta$, $\gamma$ are numerical constants.

Thus, we argue that in contrast to Born's rule (11), it is the probability of particle creation per unit time per unit volume that is proportional to $|\psi(t,\mathbf{r})|^2$ and not the probability density of finding the particle at a given point.

4. The particles are unstable formations that have a mean lifetime $\tau$ and spontaneously decay independently of each other. Thus, it is assumed that the process of particle annihilation is also a random one, and the particles are annihilated independently of the field intensity $\psi(t,\mathbf{r})$ and the presence of other particles (we do not consider here the classical electromagnetic interaction of charged particles). For this reason, the annihilation process should be considered a spontaneous process that is characterized by the mean particle lifetime

$$\tau = (\gamma\omega)^{-1} \tag{13}$$

In this case, the rate of particle annihilation $v_-$ (the probability of particle decay per unit volume per unit time) is given by a relation similar to that for radioactive decay

$$v_- = \frac{1}{\tau} p \tag{14}$$

We note that this does not violate the conservation energy law, as it is assumed that the energy required for particle birth is taken from the wave field $\psi(t,\mathbf{r})$ (the field creates the particles), and after spontaneous decay of the particles, it returns to the wave field.

Thus, for the probability density $p(t,\mathbf{r})$ of finding the particle at point $\mathbf{r}$ at time $t$, one can write a kinetic equation similar to equation (4)

$$\frac{dp}{dt} = v_+ - v_- \tag{15}$$

which when taking into account expressions (12)-(14), takes the form

$$\frac{\partial p}{\partial t} = \gamma\left(\beta \frac{|\psi(\mathbf{r},t)|^2}{\hbar} - \omega p\right) \tag{16}$$

For a given field $\psi(t,\mathbf{r})$, which can be found, for example, by solving the wave equation, the probability of finding the particle at a given point is determined by the solution of the kinetic equation (16):

$$p = \frac{\gamma\beta}{\hbar} \exp\left(-\frac{t}{\tau}\right) \int_0^t |\psi(t',\mathbf{r})|^2 \exp\left(\frac{t'}{\tau}\right) dt' \tag{17}$$

Because any measurement is always associated with averaging (due to the inertia of the devices), the instantaneous probability $p(t,\mathbf{r})$ is not of practical interest; however, its mean (observed) value, as follows, is:

$$\rho(t,\mathbf{r}) = \langle p(t,\mathbf{r}) \rangle \tag{18}$$

It is easy to calculate the mean (over time $t$) probability density of a particle, finding:

$$\langle p(t,\mathbf{r}) \rangle = \frac{\gamma \beta \tau}{\hbar} \langle |\psi(t,\mathbf{r})|^2 \rangle - \frac{\tau}{t} p(t,\mathbf{r}) \tag{19}$$

where

$$\langle |\psi(t,\mathbf{r})|^2 \rangle = \frac{1}{t} \int_0^t |\psi(t',\mathbf{r})|^2 dt' \tag{20}$$

is the mean intensity of the wave field at a given point.

Because $p(t,\mathbf{r})$ is limited at $t \to \infty$, one obtains

$$\rho(t,\mathbf{r}) = \frac{\beta}{\hbar \omega} \langle |\psi(t,\mathbf{r})|^2 \rangle \tag{21}$$

In particular, for the stationary field, we find

$$\langle |\psi(t,\mathbf{r})|^2 \rangle = |\psi(t,\mathbf{r})|^2 \tag{22}$$

A similar result is obtained for a slowly varying field $\psi(t,\mathbf{r})$.

Let the characteristic time $T$ of a change of the field $\psi(t,\mathbf{r})$ satisfy the condition

$$T \gg \tau \tag{23}$$

Then, expression (17) can be approximately written as

$$p(t,\mathbf{r}) = \frac{\beta}{\hbar \omega} |\psi(t,\mathbf{r})|^2 \tag{24}$$

Expression (24) coincides with Born's rule (11), which lies at the basis of the probabilistic interpretation of quantum mechanics. However, as we see, it is only a result of the more general principle stated above that is described by the kinetic equation (16).

For the wave function of a single particle, we use the conventional normalisation accepted in quantum mechanics

$$\int |\psi(\mathbf{r},t)|^2 d\mathbf{r} = 1 \tag{25}$$

Due to the linearity of the Schrödinger equation, the choice of normalisation of the wave function does not affect the result and leads only to a change in the constant $\beta$.

Taking into account condition (25) and the normalisation of probability density

$$\int p(\mathbf{r},t) d\mathbf{r} = 1 \tag{26}$$

we find, in the limit of (24)

$$\beta = \hbar\omega \quad (27)$$

Thus, equation (16) takes the form

$$\frac{\partial p}{\partial t} = \gamma\omega\left(|\psi(\mathbf{r},t)|^2 - p\right) \quad (28)$$

while the observable (24) is described by the expression

$$p(t,\mathbf{r}) = |\psi(t,\mathbf{r})|^2 \quad (29)$$

We see that Born's rule (11) results from our theory and follows from the exact solution (17) of the kinetic equation (28).

Let us note that, despite the fact that the proposed interpretation and traditional Copenhagen interpretation based on Born's rule (11) give nearly the same results, there is a fundamental difference between them. The Copenhagen interpretation immediately postulates a relationship between the wave function and the probability of finding the particle at a given point (11), and the particle is considered an unchanging point object, that is, it always exists, but it appears at different points in space with a different probability. It is impossible to imagine such a particle and a classical model built for it, despite many attempts that have been made [20-26].

The proposed theory considers all particles as temporary formations (field bursts, or clots) that are only perceived in the experiments as particles. These formations can be created and annihilated with some probability that is determined by the intensity of the wave field $\psi(t,\mathbf{r})$. In this interpretation, the only continuous field (wave) is a primary and physically realistic one.

As we noted in the previous section, creation and annihilation of particles are not new concepts in quantum physics, as second quantization formalism with its creation and annihilation operators initially manipulates the particles and interprets all processes as the annihilation of particles in one state and their creation in another. The new concept in our theory is that the particles are created by the wave field $\psi(\mathbf{r},t)$, and the rate of creation (12) depends on the intensity of the total wave field at a given point, which can be calculated from the wave equations, e.g., the Schrödinger equation.

By analogy with photons, this theory allows one to understand and explain the principle of indistinguishability of quantum particles and renders meaningless the concept of the particle trajectory. Assuming that a particle is a temporary formation that is created and annihilated in the wave field $\psi(\mathbf{r},t)$, we immediately come to the principle of indistinguishability of particles and lack of their trajectories.

It is easy to show that in the proposed theory, the Heisenberg uncertainty principle loses its mystical meaning regarding the fundamental limitations on the accuracy of the measurement of the particle parameters and acquires a quite classical sense. In fact, for each detection of an interaction of the particle with a device, we deal with different "particles". Each newly created particle has a momentum from some range, which is determined by the parameters of the wave field. In this case, the Heisenberg uncertainty relation only states that, the more characteristic spatial size of the wave field (the width of the wave packet), the smaller is the range of possible values that exist for the momenta of different particles created by the field and vise versa. Together with the mystical meaning of the uncertainty relations, the mystical meaning of the concept of measurement underlying the Copenhagen interpretation [7,39] disappears. We can now return to the classical notion of measurement and to the objective properties of quantum particles, independently of whether we carry out any measurements with them or not, and, in general, independently of the fact of our existence.

For the same reason, the Einstein-Podolsky-Rosen paradox [40] loses its significance, as well as Bell's theorem [41] as applied to a quantum particle as a temporary formation that does not possess a trajectory.

## IV. CONCLUDING REMARKS

In this paper, we proposed a new interpretation of the wave-particle duality and demonstrated the feasibility of its use in explaining optical and, in general, quantum phenomena. Let us summarize briefly the results of this work.

1. The proposed theory allows resolution of the conflict associated with the interpretation of the wave-particle duality.

2. We introduced a new concept of quantum particles as temporary objects permanently being created in the wave field and spontaneously decaying due to their instability. Thus, we filled, with a direct physical meaning, the standard notions of "creation" and "annihilation" of the particles used in second quantization formalism.

3. We substantiated the traditional probabilistic interpretation of quantum mechanics - Born's rule, which follows from our general concept.

4. We have shown the meaninglessness of the concept of the trajectory of quantum particles.

5. We explained the principle of indistinguishability of quantum particles.

6. We gave a natural (classical) explanation of the Heisenberg uncertainty principle: it only limits the values of momentum of the particles that can be created in the wave field and, therefore, can be detected experimentally. This removes the mystical veil from the Heisenberg uncertainty principle, which is attributed to the Copenhagen interpretation upon which it is built. For this reason, we can return to the classical notion of measurement, treating all processes as objective, regardless of whether we performed a measurement or not, to predict the quantum phenomena as such, and not treating the measurement process as a mandatory part of the physical process.

The proposed interpretation of the wave-particle duality is apparently thus far the only rational interpretation that is completely consistent with the classical physical images. It eliminates completely the conflict between the corpuscular and the wave nature of matter and allows a deeper understanding of the quantum phenomena.

Naturally, one can find similarities of this interpretation with other well-known interpretations of quantum mechanics [20-26]. One can, for example, see certain parallels between the creation-annihilation process and the collapse of the wave function considered by the Copenhagen and a number of other interpretations.

We know of no theory that is based on relations (2), (3), (5) and (12) - (14), (16) and that would argue that all quantum particles are temporary formations permanently being created in a continuous wave field and disappearing due to their spontaneous decay. For this reason, we cannot identify quantum particles (e.g., photons) with solitons because the latter are always considered long-lived field formations and therefore cannot explain wave-particle duality.

We note that the proposed interpretation not only explains (as does most of the known interpretations) the nature of photons and other quantum particles, and the nature of the wave-particle duality, it also points to possible directions for further development of quantum theory. In particular, it points to the principal possibility of building a classical field theory from which quantum mechanics follows as an approximation, focusing on the dynamics of not the field but its bursts - photons, electrons, etc. Within the limits of this concept, it is necessary to find nonlinear field equations satisfying the following conditions: (i) the approximate (linear) form of these equations must be the known wave equations (Maxwell equations, Schrödinger equation, etc.); (ii) these equations must predict multiple instabilities in a flat wave; it should disperse into a set of bursts that are interpreted in experiments as particles (e.g., photons); (iii) each burst must transport a certain energy and momentum; (iv) the bursts should be unstable temporary formations, and after some time should decay spontaneously, after which the process of creation-annihilation of new bursts is repeated; (v) the relations (2), (3), (12) - (14) and the kinetic equations (5) and (16) should follow from the field equations.

In conclusion, we note that the proposed interpretation of the wave-particle duality is of great importance for photonics in general and, in particular, for the theory of quantum computation, as an understanding of the nature of objects such as photons *allows understanding of the real abilities of quantum computing and, in general, operations with individual photons.*

# REFERENCES

[1] Einstein, A. Über einen die Erzeugung und Verwandlung des Lichtes betreffenden heuristischen Gesichtspunkt. Annalen der Physik **17** (6): 132–148 (1905).

[2] Compton, Arthur H. A Quantum Theory of the Scattering of X-Rays by Light Elements. Physical Review **21** (5): 483–502 (May 1923).

[3] Taylor, G.I. "Interference Fringes with Feeble Light", *Proc. Cam. Phil. Soc.* **15**, 114 (1909).

[4] Einstein, A. Zum gegenwärtigen Stande des Strahlungsproblems. Physikalische Zeitschrift, **10**, 185–193 (1909).

[5] Einstein, A. "Über die Entwicklung unserer Anschauungen über das Wesen und die Konstitution der Strahlung". Physikalische Zeitschrift **10**: 817–825 (1909).

[6] Pais, A. (1986). Inward Bound: Of Matter and Forces in the Physical World. Oxford University Press p. 260.

[7] Messiah, A. Quantum Mechanics. Dover Publications Inc. New York, 1999.

[8] de Broglie, L., Recherches sur la théorie des quanta, Thesis (Paris), 1924; L. de Broglie, Ann. Phys. (Paris) **3**, 22 (1925).

[9] Estermann, I.; Stern O.. "Beugung von Molekularstrahlen". Zeitschrift für Physik **61** (1-2): 95–125 (1930).

[10] Tonomura A., Endo J., Matsuda T., Kawasaki T., and Ezawa H., Demonstration of single-electron build-up of an interference pattern, Am. J. Phys. 57: 117 (1989).

[11] Doak, R.B.; R.E.Grisenti, S.Rehbein, G.Schmahl, J.P.Toennies2, and Ch. Wöll. "Towards Realization of an Atomic de Broglie Microscope: Helium Atom Focusing Using Fresnel Zone Plates". Physical Review Letters **83**: 4229–4232 (1999).

[12] Shimizu, F.. "Specular Reflection of Very Slow Metastable Neon Atoms from a Solid Surface". Physical Review Letters **86**: 987–990 (2000).

[13] Arndt, M.; Nairz, O., Voss-Andreae, J., Keller, C., van der Zouw, G., Zeilinger, A.. "Wave-particle duality of C60". Nature **401** (6754): 680–682 (14 October 1999).

[14] Gerlich, S.; S. Eibenberger, M. Tomandl, S. Nimmrichter, K. Hornberger, P. J. Fagan, J. Tüxen, M. Mayor & M. Arndt. "Quantum interference of large organic molecules". Nature Communications **2** (263) (05 April 2011).

[15] Hackermüller, L., Uttenthaler, S., Hornberger, K., Reiger, E., Brezger, B., Zeilinger, A., and Arndt, M. "The wave nature of biomolecules and fluorofullerenes". Phys. Rev. Lett. **91** (9) (2003).

[16] Couder, Y., E. Fort, Single-Particle Diffraction and Interference at a Macroscopic Scale, PRL **97**, 154101 (2006).

[17] Born, M.. "Quantenmechanik der Stoßvorgänge". Zeitschrift für Physik **38** (11–12): 803–827 (1926).

[18] Heisenberg, W., "Über den anschaulichen Inhalt der quantentheoretischen Kinematik und Mechanik", Zeitschrift für Physik **43** (3–4): 172–198 (1927).

[19] Bohr, N. Quantum postulate and recent developments in atomism. Naturwissenschaften **16**: 245-257 (1928).

[20] Tegmark M., "The Interpretation of Quantum Mechanics: Many Worlds or Many Words?", Fortschritte der Physik, **46** (6-8), 855–862 (1998) (arXiv:quant-ph/9709032v1 (1997)).

[21] Bohm, D. (1952). "A Suggested Interpretation of the Quantum Theory in Terms of "Hidden Variables" I". Physical Review **85**: 166–179.

[22] Bohm, D. (1952). "A Suggested Interpretation of the Quantum Theory in Terms of "Hidden Variables", II". Physical Review **85**: 180–193.

[23] Ballentine, L. E. The Statistical Interpretation of Quantum Mechanics, Rev. Mod. Phys. (1970) **42**, 358–381.

[24] Cramer, J. G. The transactional interpretation of quantum mechanics, Rev. Mod. Phys. (1986) **58**, 647–687.

[25] Omnès, R. Consistent interpretations of quantum mechanics, Rev. Mod. Phys. (1992) **64,** 339–382.

[26] Schlosshauer, M. Decoherence, the measurement problem, and interpretations of quantum mechanics, Rev. Mod. Phys. (2005) **76**, 1267–1305.

[27] Feynman, R. P.. *The Feynman Lectures on Physics, Vol. 3*. USA: Addison-Wesley. (1965).

[28] Wiener. O. Stehende Lichtwellen und die Schwingungsrichtung polarisirten Lichtes. Annalen der Physik und Chemie **40**, 203-243 (1890).

[29] P. Drude, W. Nernst, Wiedem. Ann **45**, 460 (1892).

[30] H. E. Ives, T. C. Fry, J. Opt. Soc. Amer. **23**, 73 (1933).

[31] Born, M., and Wolf, E., Principles of Optics: Electromagnetic Theory of Propagation, Interference and Diffraction of Light, Cambridge University Press, 1999.

[32] Berestetskii, V.B., Lifshitz, E.M., Pitaevskii, L.P. (1982). *Quantum Electrodynamics*. **Vol. 4** (2nd ed.). Butterworth-Heinemann.

[33] Braun, A., Korn, G., Liu, X., Du, D., Squier, J., Mourou, G., (1995). Self-channeling of high-peak-power femtosecond laser pulses in air. Opt. Lett. 20 (1), 73–75.

[34] Couairon, A*.,* and Mysyrowicz, A. Femtosecond filamentation in transparent media. Physics Reports 441 (2007) 47 – 189.

[35] Bergé, L., Skupin, S., Nuter, R., Kasparian, J., and Wolf, J.-P. Ultrashort filaments of light in weakly ionized, optically transparent media. (2007) *Rep. Prog. Phys.* **70** 1633.

[36] Zurek, W. H. Environment-Assisted Invariance, Entanglement, and Probabilities in Quantum Physics. 2003 *Phys. Rev. Lett.* **90** 120404.

[37] Zurek, W. H. Probabilities from entanglement, Born's rule $p_k = |\psi_k|^2$ from envariance. Phys Rev A 71, 052105 (2005).

[38] Schlosshauer, M., and Fine, A. On Zurek's Derivation of the Born Rule. Foundations of Physics, **35** (2), 197-213, (2005).

[39] J. A. Wheeler and W. H. Zurek, *Quantum Theory and Measurement* (Princeton University Press, Princeton, NJ, 1983).

[40] Einstein, A., Podolsky, B. Rosen, N.: Can quantum-mechanical description of physical reality be considered complete?. Phys. Rev. **47** (10), 777780 (1935).

[41] Bell, J. Speakable and Unspeakable in Quantum Mechanics. Cambridge Univ. Press, Cambridge (1987).